\documentclass[12pt,a4paper]{article}
\usepackage{amsmath}
\usepackage{subcaption}
\usepackage{amssymb}
\usepackage[numbers]{natbib}
\usepackage{authblk}
\usepackage{epsfig}
\usepackage{graphicx}
\usepackage{float}
\usepackage{IEEEtrantools}
\usepackage{indentfirst}

\begin{document}
	\title{Exit dynamics from Morse potential \\under thermal fluctuations}
	\author{Vipin. P, R. Sankaranarayanan}
	\affil{Department of Physics\\ National Institute of Technology, Tiruchirappalli-620015\\ Tamilnadu, India.}
	\date{}
	\maketitle{}
	
	\begin{abstract}
		We study the dynamics of a Brownian particle in Morse potential under 
		thermal fluctuations, modeled by Gaussian white noise whose amplitude depends on absolute temperature. Dynamics of such a particle is investigated by numerically integrating the corresponding Langevin equation. From the mean first passage time (escape time), we study the dependence of Kramer's rate on temperature and viscosity of medium. An approximate analytical expression for the rate constant is found by solving differential equation for the mean first passage time. The expression shows a temperature dependent pre-factor for the Arrhenius equation. Our numerical simulations are in agreement with the analytical approximations.
	\end{abstract}
	
	\section*{Introduction}
	Chemical kinetics is an important subsection of physical chemistry. It deals with the rate of chemical reaction and its dependence on concentration of reactants/products and on temperature. The most widely used equation that describes the temperature dependence of reaction rate constant is the Arrhenius equation, which is proposed by Svante Arrhenius in the year 1889. In fact, Arrhenius equation is one of the three solutions of a first order differential equation which was originally proposed by Jacobus Henricus van't Hoff in the year 1884 \citep{laidler_chemical_1985}. The importance of Arrhenius' work in this regard is that he adopted and applied his solutions to a wide variety of chemical reactions. van't Hoff derived his famous equation from pure thermodynamical considerations on a mixture of ideal gases. \par Another approach to the calculation of reaction rates is the theory of fluctuations where the reaction rates are calculated as rate of escape from metastable states, as pioneered by Farkas \citep{farkas1927}. A major breakthrough in this direction came in 1940 when Kramers, in his seminal work, showed that the van't Hoff-Arrhenius equation can be obtained by solving a Smoluchowsky equation in the strong friction limit \citep{kramers1940brownian}. Kramers considered the escape dynamics of a Brownian particle trapped inside an asymmetric double well potential which is acted upon by a Gaussian white noise. Since then escape from metastable states have been an active area of research in the area of chemical physics and physical chemistry and still remains as one of the major approaches in the theory of reaction rates. \par In this work, we consider the motion of a classical Brownian particle inside a Morse potential. Though the Morse potential was originally introduced and subsequently used predominantly for quantum mechanical calculations of vibrational spectra and dissociation of diatomic molecules, it has also been shown that the potential can have classical solutions corresponding to both confined and dissociated states  \citep{barboza2007analytical, demarcus1978classical,slater1957classical}.
	Also, in a classical setting, the Morse potential have been used to model the interaction between two pairs of bases in DNA \citep{peyrard2004nonlinear} and in the theory of collision induced emission (CIE) from a pair of dissimilar atoms \citep{reguera2006classical}.The temperature dependence of the escape from the potential well can be extracted from the diffusion term of the Langevin equation that governs the dynamics of the Brownian particle. In the context of chemical reactions, our study attempts to explore the classical dissociation of a diatomic molecule under thermal fluctuations.
	\par In the framework of activated rate processes, the reaction rate is calculated from the relation \citep{hanggi1990reaction,dybiec2007escape}
	\begin{equation}
	\kappa = \frac{1}{\tau}
	\end{equation}
	where $\tau$ is the \textit{mean first passage time\hspace*{1mm}(MFPT)} for the particle in the well. The mean first passage time for a stochastic process in a bounded domain is defined as the average time elapsed before the process, which started at one of the local minima of the domain, had exited the domain for the first time \citep{gardiner1985stochastic}.
	Usually, the parabolic top of the potential function is chosen as the boundary  (transition point). In the language of transition state theory, this point corresponds to a transition state \citep{henriksen2018theories}. Once the reaction coordinate crosses this point from the reactant state, the reaction is said to have occurred in the forward direction and the particle is said to have escaped from the well. The inverse of this mean first passage time is defined as the rate of the reaction. For double well potential in which there is equal probability to escape in either direction, the total escape time would be $2 \tau$ and the escape rate would be one half of that given by equation (1).
	\section*{First passage time}
	
	The Morse potential was proposed by Phillip M. Morse in 1929, in order to describe the vibrational anharmonicity of diatomic molecules. It is the natural choice to describe dissociation reactions of diatomic molecules. A schematic diagram  of the potential considered is shown in Fig.1.
	\begin{figure*}
		\begin{center}
			\centering		\includegraphics*[width=0.75\linewidth]{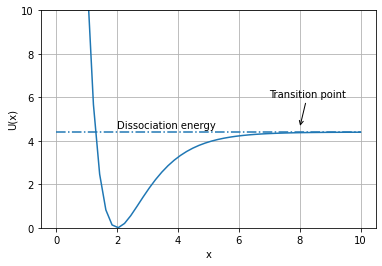}
			\caption{\small The Morse potential with $U_{0} = 4.4$, $g = 0.98$. }
		\end{center}
	\end{figure*}
	The Morse potential is given by the expression
	\begin{equation}
	U(x)=U_0(1-e^{-g x})^2
	\end{equation}0
	where $U_0$ is the depth of the well, $x$ is the momentary displacement of the bond length from its equilibrium length and $g$ is a shape parameter which is given by
	\begin{equation}
	g = \sqrt{\frac{k}{2U_0}}
	\end{equation}
	with $k$ being the force constant of the molecule in equilibrium. When $x=0$, $U(x)= 0$, the molecule is at the bottom of the well and is in equilibrium.
	\par We consider a particle of unit mass undergoing a dissociation reaction in an inert medium of viscosity $\gamma$. The dynamics of such a system can be described by Langevin equation for a dynamic quantity associated with the reaction. Such a quantity which represents the physical system in the equation is called a reaction coordinate . According to the \emph{Gold book} of IUPAC \citep{gold1997}, a reaction coordinate is defined as a geometric parameter that changes during the conversion of one\hspace{1mm}(or more) reactant molecular entities into one \space (or more) product molecular entities and whose value can be taken for the measure of the progress of an elementary reaction. Thus, for a dissociation reaction, the best choice for the reaction coordinate is the momentary displacement of  the bond length of the diatomic species.  
	The equation of motion for such a reaction is given by
	\begin{equation}
	\dot{x}=-\frac{{U}\sp\prime(x)}{\gamma}+\alpha \xi(t).
	\end{equation}
	Here $\alpha$ is a parameter that regulates the amplitude of the noise $\xi(t)$. It is related to absolute temperature $T$ and coefficient of viscosity $\gamma$ through the relation
	\begin{equation}
	\alpha = \frac{k_BT}{\gamma}
	\end{equation} 
	where $k_B$ is the Boltzmann constant and we set $k_B=1$ . We consider a Gaussian distributed random variable (noise) with properties
	\begin{equation}
	\langle \xi(t) \rangle = 0 , \qquad  \langle \xi(t)\xi(t+s) \rangle = 2k_BT\delta(t-s).
	\end{equation}
	When $\alpha = 0$, the process is not activated and the particle will remain in the well for infinitely long time. As the strength of the fluctuations begins to build up, the random force on the particle begins to accumulate and eventually the particle will escape the well.
	
 	The Langevin equation is solved using the Euler-Maruyama method \citep{rossant2018ipython, capala2020evy}. According to the scheme, eq. $(4)$ is discretized as
	\begin{equation}
	x_{i+1} = x_i - \frac{h}{\gamma}U'(x_i)\Delta t+\alpha \xi(t) (\Delta t)^{1/2}
	\end{equation}
	where $x_i$ stands for $x(t_{i})$, $h$ is the time step and $\xi(t)$ is a random variable drawn from the normal distribution $\mathcal{N} (0,1)$. The term $(\Delta t)^{(1/2)}$ comes from the quadratic variation of the Wiener process. Since error at each step of the integration is $O[h]$ the scheme is best suited of stochastic differential equations with additive noise \citep{toral2014}. The sample trajectories of the particle are shown in Fig.\hspace{1mm}2 for different values of $\alpha$. As expected, the particle tends to escape the well faster as $\alpha$ increases.
	\begin{figure}[!t]
			\centering
			\begin{subfigure}{0.75\textwidth}
					\caption{}
			\includegraphics*[width=\textwidth]{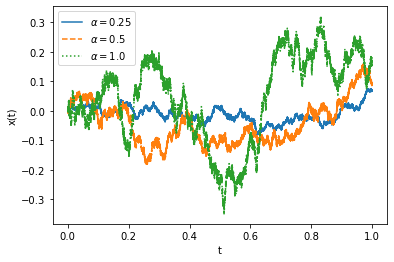}\\
		
			\end{subfigure}
			\begin{subfigure}{0.75\textwidth}
				\caption{}
			\includegraphics*[width=\textwidth]{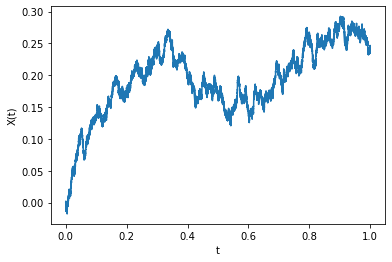}
			
			\end{subfigure}
			\caption{\small (Colour online)\hspace{1mm}(a) Sample trajectories for different noise amplitudes with $U_{0} =4.4$ and $g = 0.98$. (b) Trajectory averaged over $10^{4}$ realizations for $\alpha = 0.25 $. }
	
	\end{figure}
	
	In the context of chemical reactions, first passage time is the time needed for a reaction to occur for the first time. Since the collisions of molecules are random, the first passage time itself is a random quantity. All the collisions of the reactant molecule with the inert medium will not result in a reaction. 
	The first passage time is calculated by numerically integrating the Langevin equation by imposing absorbing boundary condition at the transition point. This is achieved by stopping the integration once the particle coordinate crosses the transition point. The corresponding time step is recorded and the particle is reinstated to the bottom of the well. The time step $h = 10^{-5}$ and the first passage time is averaged over $10^4$ realizations. We may note here that in the case of potentials with a maxima, the transition point is well defined. But in the case of Morse potential and other similar potentials like Lennard-Jones, such a maxima does not exist. In these kind of potentials, escape can only be defined arbitrarily as the attainment of certain distance from the well \citep{larson1988thermally}.
	\begin{figure*}
		\begin{center}
			\centering		\includegraphics*[width=0.75\linewidth]{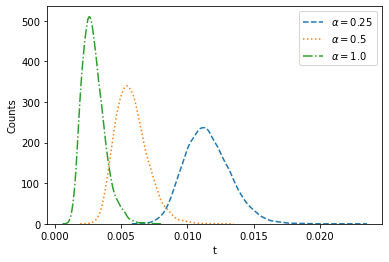}
			\caption{\small Distribution of first passage times for $U_{0} = 4.4$, $g = 0.98$. The absorbing boundary is set at $x = 8$ where the potential function is nearly flat, as shown in figure 1.}
		\end{center}
	\end{figure*}
	Fig. 3 shows the distribution of first passage times for different values of $\alpha$. The distributions are unimodal, and the peak shift towards left for larger noise amplitudes as we expect from intuition.

	From the collected data of first passage times, one can also estimate the cumulative distribution of the first passage times. In the context of chemical reactions, this is called the dissociation probability. Dissociation probability is the probability for the diatomic molecule to dissociate. Kenfack and Rost \citep{kenfack2005stochastic} studied the stochastic dissociation of diatomic molecules by solving a Langevin equation similar to eq. (4) for Morse potential. The numerical study was carried out under the action of a random force modeled by a white shot noise of the form
	\begin{equation}
	F_{i}(t) = \sum_{i=1}^{N_i}d_i\delta(t-t_i)
	\end{equation}
	where $F_i$ is the random force and $d_i$ is the strength of the force. They found that the classical dissociation probability can be parametrized with the expression
	\begin{equation}
	P_{d}(t) = \frac{1}{2} \tanh(at+b)+\frac{1}{2}
	\end{equation}  
	where $a$ and $b$ are constants. We find that the dissociation of diatomic molecule under Gaussian noise fits quite well with the above expression as shown in Fig. 4(a). With this one can define the complimentary function, survival probability
	\begin{equation}
		S(t) = 1-P_{d}(t).
	\end{equation}
	
\begin{figure}[!t]
			\centering
			\begin{subfigure}{0.75\textwidth}
				\caption{}
			\includegraphics*[width=\textwidth]{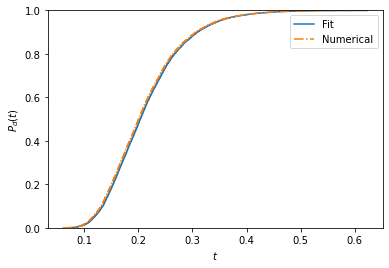}\\
			
			\end{subfigure}
			\begin{subfigure}{0.75\textwidth}
				\caption{}
			\includegraphics*[width=\textwidth]{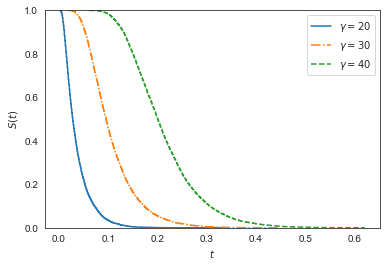}

			\end{subfigure}
		\caption{\small (Colour online)\hspace{1mm}(a) Dissociation probability. The values chosen for the parameters are $U_0 = 4.4$, $\gamma = 40$, $a = 0.98$ and $T = 250$.   (b) Survival probability for various values of viscosity parameter. Other parameter values are as mentioned before. }
		
	\end{figure}
	 
	 We must note here that, unlike in the Kramer's reaction rate theory where the survival probability decay exponentially \citep{metzler2000kramers}, decays non-exponentially as shown in the Fig. 4(b). Such a non-exponential decay is a regular feature of diffusion controlled dissociation reactions \citep{berg1978diffusion}.  The decay has also been reported in escape of a Brownian particle from a double Morse potential \citep{shizgal2015spectral} whose dynamics obeys Fokker-Planck equation. In addition, we also calculate the survival probability for different values of the viscosity of the medium as shown in Fig. 4(b). The plots suggest that larger the viscosity of the medium, longer the particle dwell within the well region. For smaller viscosity, the survival probability falls steeply, indicating a faster exit from the well and so is the acceleration in dissociation. 
	\subsection*{Reaction rate}
	 As mentioned in the introduction, the rate of escape from the potential well can be computed from the mean first passage time. In order to estimate the mean first passage time, the Langevin equation is integrated numerically by setting up absorbing boundary at the transition point. The first passage time is computed by stopping the integration once the reaction coordinate crosses the transition point and recording the corresponding time. For given values of $\gamma$ and $T$, the process is repeated for $10^4$ realizations and the average of these first passage times is calculated. The reaction rate is computed for different values of $T$ and the data are plotted.  While Fig. 5(a) shows the behaviour of reaction rate as a function of $T$ for a given value of $\gamma$, Fig. 5(b) depicts the dependence of reaction rate on coefficient of viscosity $\gamma$ for a given value of $T$.
	 
	 The numerical data in Fig. 5(a) fits well with a straight line of slope $\approx 0.522$, in log-log scale, implying that the reaction rate $\kappa \propto \sqrt{T}$. On the other hand, Kramers found that the reaction rate constant is inversely proportional to the viscosity, at a given temperature $T$,  through the pre-factor given by
	\begin{equation}
	A_{\kappa} = \frac{\omega_a \omega_b}{2\pi \gamma}
	\end{equation}
	where $\omega_a$ and $\omega_b$ are harmonic frequencies associated to the minimum and maximum of the double well potential respectively. Fig. 5(b) shows the variation of rate constant with respect to the viscosity of the medium. The data fits well with a straight line of slope $\approx -0.99$ in log-log scale, indicating that $\kappa \propto 1/\gamma$ as in the case of Kramers rate theory. This agrees well with the survival probability studies with different values of $\gamma$ as shown earlier.
\begin{figure}[!t]
			\centering
			\begin{subfigure}{0.75\textwidth}
			\caption{}
			\includegraphics*[width=\textwidth]{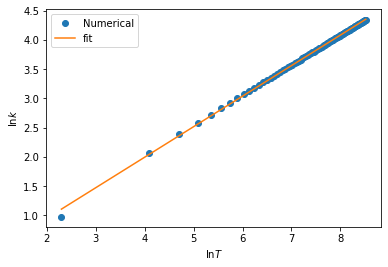}\\
			
			\end{subfigure}
			\begin{subfigure}{0.75\textwidth}
			\caption{}
			\includegraphics*[width=\textwidth]{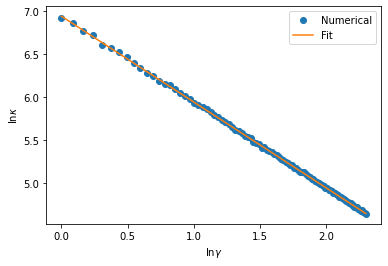}
			\end{subfigure}
	
		\caption{\small (Colour online) (a) Reaction rate versus temperature for $\gamma = 20$. The slope of straight line is obtained as $0.522$. (b) Reaction rate versus $\gamma$ at $T = 500 $. The numerical data is fitted with a straight line of slope $-0.99$. }
	
	\end{figure}  
	\par In general, the temperature dependence of dissociation rate constant can have the form
	\begin{equation}
	\kappa(T) = {\cal{A}}T^{m}\exp\left(-E_a/RT\right)
	\end{equation}
	where $E_a$ is the activation energy and $R$ is the universal gas constant.To find $m$, we start by writing the differential equation for the mean first passage time $\tau$ as [10]
	\begin{equation}
	A(x)\frac{d\tau}{dx}+D\frac{d^{2}\tau}{dx^{2}} = -1
	\end{equation}
	where $A(x) = U'(x)/\gamma$ and $D = \alpha/2$. Multiplying the above equation with an integration factor $e^{-\beta U}$ where $\beta = 1/k_{B}T$ and integrating twice, we get the general solution as
	\begin{equation}
	\tau = \frac{1}{D}\int_{a}^{b}e^{\beta U(y)}dy\int_{-\infty}^{y}e^{-\beta U(z)}dz
	\end{equation}
	where $a$ corresponds to the bottom of the potential well, $x$ is an arbitrary point inside the well beyond $a$ and $b$ is the transition point. 
	As we move from $a$ to $b$, $U(y)$ increases and $e^{\beta U(y)}$ increases even faster. Hence the integrand of the first integral is dominated around the transition point $b$, where the potential is nearly constant. Expanding the potential in Taylor series about this point, we have
	\begin{equation}
	U(y) = U(b)+U'(b)(y-b)+\cdots
	\end{equation}
	Hence the first integral can be approximated as
	\begin{equation}
	\int_{a}^{b}e^{\beta U(y)}dy \approx \int_{a}^{b}e^{\beta U_0}dy
	\end{equation}
	where $U(b) = U_0$, the dissociation energy. Evaluating the integral, we get
	\begin{equation}
	\int_{a}^{b}e^{\beta U_0}dy = e^{\beta U_0}(b-a).
	\end{equation}
	Note that the integral depends on the exact location of the transition point, in contrast to the models with a parabolic barrier top.
	
	The second integral is maximum at the bottom of the well where we can linearize the potential as
	\begin{equation}
	U(z) \approx U(a)+\frac{1}{2}m \omega^{2}(z-a)^2
	\end{equation}
	such that $\omega = \sqrt{U''(a)/m}$ is the frequency inside the well. Thus the second integral becomes
	\begin{equation}
	\int_{-\infty}^{y}e^{-\beta U(z)}dz \approx e^{-\beta U(a)}\int_{-\infty}^{+\infty}e^{-\frac{\beta}{2}m{\omega}^2(z-a)^2}dz = e^{-\beta U(a)} \sqrt{\frac{2 \pi}{\beta m{\omega}^2}}.
	\end{equation}
	Note that this is a Laplace integral. This means that the contribution beyond a small region near the bottom is subdominant \citep{bender2013}. Thus the upper limit can safely be set to $+\infty$. 
	Substituting equations $(16)$ and $(18)$ in $(14)$ we get
	\begin{equation}
	\tau = \sqrt{\frac{2 \pi \beta}{m}}\frac{2\gamma(b-a)}{\omega}  e^{\beta[ U_0- U(a)]}.
	\end{equation}
	Thus the rate is given by
	\begin{equation}
	\kappa = \frac{1}{\tau}= \frac{\omega}{2\gamma (b-a)} \sqrt{\frac{m}{2 \pi \beta}}e^{-\beta[ U_0- U(a)]}.
	\end{equation}
	Referring $U_0- U(a) = E_a$, the activation energy and setting
	\begin{equation}
	\omega\sqrt{\frac{m k_B}{8\pi}}= \cal{A}
	\end{equation}
	we get
	\begin{equation}
	\kappa = {\cal{A}}\frac{\sqrt{T}}{\gamma (b-a)}e^{-Ea/k_BT}.
	\end{equation}	
	Thus we observe that the reaction rate is proportional to $\sqrt{T}$ and inversely proportional to viscosity $\gamma$. This is indeed the case as shown from the numerical results.
	\subsection*{Conclusion}
	In order to model diatomic dissociation in inert media, we have analyzed the exit dynamics of a Brownian particle from Morse potential under thermal fluctuations characterized by Gaussian white noise. Unlike other chemical reactions modeled by double well potential, probability of the particle staying inside the bottom of the well shows non-exponential decay with time. Our analytical study reveals pre-factor for the reaction rate constant which depends on temperature and viscosity of the media. The same is being confirmed through numerical solutions of the corresponding Langevin equation. In addition, the first passage time is found to be directly proportional to the location of the transition point. This is intuitive because, beyond the transition point, particle is force-free and hence move with a constant velocity. Thus the time to reach the transition point is directly proportional to the distance.

\end{document}